\documentclass[3p]{elsarticle}
\usepackage{epsf,epsfig,graphicx}
\usepackage{amsmath}
\usepackage{ulem}
\usepackage{color}
\setcitestyle{square}
\biboptions{sort&compress}

\begin{document}

\title{Factorization for Substructures of Boosted Higgs Jets\tnoteref{t1}}

\author{Joshua Isaacson}
\ead{isaacs21@msu.edu}
\address{Dept. of Physics \& Astronomy,
Michigan State University, E. Lansing, MI 48824, USA}

\author{Hsiang-nan Li}
\ead{hnli@phys.sinica.edu.tw}
\address{Institute of Physics,
Academia Sinica, Taipei, Taiwan 115, Republic of China,}

\author{Zhao Li}
\ead{zhaoli@ihep.ac.cn}
\address{Institute of High Energy
Physics, Chinese Academy of Sciences, Beijing 100049, China}

\author{C.-P. Yuan}
\ead{yuan@pa.msu.edu}
\address{Dept. of Physics \& Astronomy,
Michigan State University, E. Lansing, MI 48824, USA}

\tnotetext[t1]{This article is registered under preprint number:  arXiv:1505.06368}


\begin{abstract}

We present a perturbative QCD factorization formula for substructures
of an energetic Higgs jet, taking the energy profile resulting from
the $H\to b\bar b$ decay as an example. The formula is written as a
convolution of a hard Higgs decay kernel with two $b$-quark jet functions
and a soft function that links the colors of the two $b$ quarks.
We derive an analytical expression to approximate the energy
profile within a boosted Higgs jet, which significantly differs from those of
ordinary QCD jets. This formalism also extends to boosted $W$ and $Z$ bosons
in their hadronic decay modes, allowing an easy and efficient discrimination
of fat jets produced from different processes.

\end{abstract}
\maketitle
%
%

The Higgs boson, which is responsible for the electroweak
symmetry breaking mechanism in the Standard Model (SM), has been
discovered at the Large Hadron Collider (LHC) with its mass around 125 GeV.
Though its couplings to other particles seem to be consistent with the
SM, the ultimate test as to whether this observed particle is
the SM Higgs boson relies on the measurement of the trilinear
Higgs coupling that appears in Higgs pair production.
A Higgs boson is predominantly produced at rest via gluon fusion processes 
at the LHC. It has been
shown~\cite{deFlorian:2013jea} that the cross section of the Higgs pair
production increases rapidly with center-of-mass energy of
hadron colliders. With much higher collision energy in the
partonic process, preferred for exploring the trilinear Higgs coupling,
the Higgs boson and its decay products will be boosted.
An energetic Higgs boson
can also be associately produced with other SM particles,
such as $W$, $Z$ bosons, top quarks and jets~\cite{Butterworth:2008iy}.

The SM Higgs boson decays into a pair of bottom quark and antiquark dominantly.
When the Higgs boson is highly boosted, this
pair of bottom quarks may appear as a single jet and cannot be
unambiguously discriminated from an ordinary QCD jet.
A similar challenge applies to the
identification of other boosted heavy particles, e.g., $W$ bosons,
$Z$ bosons, and top quarks, when decaying via hadronic modes.
Hence, additional information on internal structures of these
boosted jets (such as their masses, energy profiles, and
configurations of subjets) is required for the experimental identification.
Many theoretical efforts were
devoted to the exploration of heavy particle jet properties
based on event generators~\cite{Almeida:2008yp,Almeida:2010pa,Thaler:2010tr}.
Recently, the perturbative QCD
(pQCD) formalism, including fixed-order calculations \cite{Almeida:2008tp} and
the resummation technique~\cite{Li:2011hy}, was employed to investigate jet
substructures. The alternative approach
based on the soft-collinear effective theory and its application to
jet production at an electron-positron collider were presented in
Refs.~\cite{Ellis:2010rwa,Kelley:2011tj}.

In this Letter we develop a pQCD factorization formula
to describe the internal jet energy profile (JEP) of the boosted
jet resulting from the $H\to b\bar b$ decay,
with energy $E_{J_H}$ and invariant mass $m_{J_H}$.
The basic idea of our theoretical approach is as follows.
A Higgs boson is a colorless particle, while its decay products,
the bottom quark and antiquark, are colored objects and
dressed by multiple gluon radiations to form a system with
mass of ${\mathcal O}(m_{J_H})$ and energy of ${\mathcal O}(E_{J_H})$.
The invariant mass $m_J$ of the bottom quark and its
collimated gluons, with energy of ${\mathcal O}(E_{J_H})$,
typically satisfies the hierarchy $E_{J_H}\gg m_{J_H} \gg m_J$.
Based on the factorization theorem,
QCD dynamics characterized by different scales must factorize
into soft, collinear, and hard pieces, separately.
First, the Higgs jet function $J_H$ is factorized from a
Higgs boson production process at the leading power of $m_{J_H}/E_{J_H}$.
Then the $b$-quark jet function is defined at the leading
power of $m_J/m_{J_H}$~\cite{Li:2011hy},
soft gluons with energy of ${\mathcal O}(m_{J_H})$
are absorbed into a soft function $S$, and the remaining energetic gluons with
energy ${\mathcal O}(E_{J_H})$ and invariant mass of ${\mathcal O}(m_{J_H})$
go into the hard Higgs decay kernel $H$.

The Higgs JEP is then factorized at leading
power of $m_J/m_{J_H}$ into a convolution of the hard kernel with two $b$-quark
jet functions and a soft function that links colors of the two $b$ quarks.
We will demonstrate a simple scheme, in which the soft gluons
are absorbed into one of the $b$-quark jets, forming a fat jet, and the
soft function reduces to unity. The other $b$-quark jet is a thin jet to avoid
double counting of soft radiation.
Evaluating the decay $H\to b\bar b$ up to leading order (LO) in
the coupling constant $\alpha_s$ and substituting the light-quark jet function
in~\cite{Li:2011hy} for the $b$-quark jet functions, we predict the Higgs JEP.
Since a Higgs boson is massive and a color singlet, its JEP
dramatically differs from that of ordinary QCD jets.
Below, we present the derivation of the JEP of a
Higgs boson decaying into a bottom-quark pair.

The four-momentum of the Higgs jet can be written as
$P_{J_H}=E_{J_H}(1,\beta_{J_H},0,0)$, with
$\beta_{J_H}=\sqrt{1-(m_{J_H}/E_{J_H})^2}$.
We define the Higgs jet function at the scale $\mu$ as
\begin{eqnarray}
J_H(m_{J_H}^2,E_{J_H},R,\mu^2)=\frac{2(2\pi)^3} {
E_{J_H}}\sum_{N_J} \langle 0|\phi(0) |N_J\rangle\langle N_J|
\phi^\dagger(0)|0\rangle\\
\times\delta(m_{J_H}^2-\hat m_{J_H}^2(N_J,R))\delta(E_{J_H}-E(N_J))
\delta^{(2)}(\hat n_{J_H}-\hat n(N_J)),\nonumber
\end{eqnarray}
where the coefficient has been chosen to satisfy
$J_H^{(0)}=\delta(m_{J_H}^2-m_H^2)$
at the zeroth order in the Yukawa coupling. $m_H$ represents the
Higgs boson mass, and $R$ the Higgs jet cone radius.
The three $\delta$-functions in the above definition specify
the Higgs jet invariant mass, energy, and unit momentum direction of the set
$N_J$ of final-state particles, respectively.
After applying the aforementioned factorization procedure, $J_H$ is written as
\begin{align}
&J_{H}(m_{J_H}^2,E_{J_H},R,\mu^2) = \frac{1}{E_{J_H}}\Pi_{i=1,2}\int
dm_{J_i}^2dE_{J_i}d^2\hat n_{J_i}\int d\omega S(\omega,R,\mu_f^2)
J_i(m_{J_i}^2,E_{J_i},R_i,\mu_f^2) \nonumber \\
 \times&H(P_{J_1},P_{J_2},R,\mu^2,\mu_f^2)
\delta(m_{J_H}^2-P_{J_1}\cdot P_{J_H}-P_{J_2}\cdot P_{J_H}-\omega)
\times\delta(E_{J_H}-E_{J_1}-E_{J_2})
\delta^{(2)}(\hat n_{J_H}-\hat n_{J_1+J_2}),\label{hj1}
\end{align}
where the factorization scale $\mu_f$ is introduced by the $b$-quark jet
functions $J_i$. $m_{J_i}$ ($E_{J_i}$, $P_{J_i}$, $R_i$) is the invariant mass
(energy, momentum, radius) of the $b$-quark jets, and the soft function takes
the form $S^{(0)}=\delta(\omega)$ at LO with the variable
$\omega\equiv P_S\cdot P_{J_H}$, where $P_S$ is the soft gluon momentum.

To describe the Higgs JEP, we define the jet energy function
$J_{H}^E(m_{J_H}^2,E_{J_H},R,r,\mu^2)$ by including
in Eq.~(\ref{hj1}) a step
function $\Theta(r-\theta_j)$ for every final-state particle $j$.
The final-state particles with non-vanishing step functions
(i.e., emitted within the test cone of radius $r$) and
associated with the $b$-quark jet $J_{i}$
are grouped into the $b$-quark jet energy function
$J_i^E(m_{J_i}^2,E_{J_i},R_i,r_i,\mu_f^2)$.
The energetic final-state particles outside the $b$-quark jets
and within the test cone are absorbed into the hard kernel $H^E$.
The other final-state particles outside the test cone
are absorbed back into their original functions.
In this work, we will consider only the LO hard kernel, for which
$H^E =H^{(0)}$. We then arrive at
\begin{eqnarray}
& &J_{H}^E(m_{J_H}^2,E_{J_H},R,r,\mu^2) =\frac{1}{E_{J_H}}\Pi_{i=1,2}\int
dm_{J_i}^2dE_{J_i}d^2\hat n_{J_i}\int d\omega S(\omega,R,\mu_f^2)
\nonumber\\
& &\times
\sum_{i\not= j}J_i^E(m_{J_i}^2,E_{J_i},R_i,r_i,\mu_f^2)
J_j(m_{J_j}^2,E_{J_j},R_j,\mu_f^2)
H^{(0)}(P_{J_1},P_{J_2},R,\mu^2,\mu_f^2)\nonumber\\
& &\times
\delta(m_{J_H}^2-P_{J_1}\cdot P_{J_H}-P_{J_2}\cdot P_{J_H}-\omega)
\delta(E_{J_H}-E_{J_1}-E_{J_2})
\delta^{(2)}(\hat n_{J_H}-\hat n_{J_1+J_2}),\label{he0}
\end{eqnarray}
where the LO hard kernel is
\begin{eqnarray}
H^{(0)}= \frac{N_c}{2\pi^3}\left(\frac{m_b}{v}\right)^2\frac{(E_{J_1}E_{J_2})^2[1-\cos(\theta_{J_1}+\theta_{J_2})]}
{(P_{J_H}^2-m_H^2)^2+\Gamma_H^2 m_H^2},
\label{Hfunction}
\end{eqnarray}
with the number of colors $N_c$, the $b$-quark mass $m_b$, the vacuum expectation
value $v$, the Higgs decay width $\Gamma_H$, and the polar angle $\theta_{J_i}$
of the $b$-quark jet $J_i$ relative to the Higgs jet axis.

We have the freedom to choose the jet parameters $R_i$ and $r_i$,
whose values depend on the scheme adopted to factorize the soft radiation
in the Higgs jet into different convolution pieces in Eq.~(\ref{he0}).
A simple scheme is to take $J_1$ as a thin jet, such that its entire energy is counted when
a sufficient amount of the thin jet is within the test cone as specified below.
For that, we set $R_1=r_1=r$, which increases from the minimal value 0.1 in our numerical
analysis. This choice leads to the simplification of $J_1^E$,
$J_1^E(m_{J_1}^2,E_{J_1},r,r,\mu_f^2)
=E_{J_1}J_1(m_{J_1}^2,E_{J_1},r,\mu_f^2)\Theta(ar-\theta_{J_1})$, with
$a\sim {\cal O}(1)$ being a geometric factor.
The scheme also includes a fat jet $J_2$ with a large cone radius $R_2=R$,
which then absorbs all soft radiation in the Higgs jet.
The energy function of the fat jet $J_2^E(m_{J_2}^2,E_{J_2},R,r_2=r,\mu_f^2)$,
which contributes to the Higgs JEP as $\theta_{J_2}\le ar$,
will take the result derived in the resummation technique~\cite{Li:2011hy,Li:2012bw}.

Next, we integrate out the dependence on the Higgs jet invariant mass by taking
the first moment in the Mellin transformation of $J_{H}^E$, defined as
${\bar J}_{H}^E(1,E_{J_H},R,r,\mu^2)\equiv
\int J_{H}^E(m_{J_H}^2,E_{J_H},R,r,\mu^2) d m_{J_H}^2/(R E_{J_H})^2$.
To perform the integration over $\hat n_{J_2}$, we write the
corresponding $\delta$-function as
\begin{eqnarray}
\delta^{(2)}(\hat n_{J_H}-\hat n_{J_1+J_2})&=&
\delta\left(\frac{{\bf P}_{J_H}}{|{\bf P}_{J_H}|}-\frac{{\bf
P}_{J_1}+{\bf P}_{J_2}}{|{\bf P}_{J_1}+{\bf P}_{J_2}|}\right),\nonumber\\
&=&\frac{|{\bf P}_{J_1}+{\bf P}_{J_2}|}{|{\bf
P}_{J_2}|}\delta\left(\frac{|{\bf P}_{J_1}+{\bf P}_{J_2}|}{|{\bf
P}_{J_H}||{\bf P}_{J_2}|}{\bf P}_{J_H}-\frac{{\bf P}_{J_1}}{|{\bf
P}_{J_2}|}-\hat n_{J_2}\right),
\end{eqnarray}
where the ratio is given by $|{\bf P}_{J_1}+{\bf P}_{J_2}|/|{\bf P}_{J_2}|=
(E_{J_1}\cos\theta_{J_1}+E_{J_2}\cos\theta_{J_2})/E_{J_2}$.
The angular relation
$E_{J_1}\sin\theta_{J_1}=E_{J_2}\sin\theta_{J_2}$ is then demanded.
The integration over $E_{J_2}$ is trivial.
The $b$-quark jet masses $m_{J_i}$, whose typical values are much
lower than $m_H$, are negligible in the hard kernel.
The integrations over $m_{J_H}^2$ and
$m_{J_i}^2$ can then be done trivially,
with $\int dm_{J_i}^2 J_i(m_{J_i}^2,E_{J_i},R_i,\mu_f^2)=1+O(\alpha_s)\approx 1$.

The soft function is defined as a vacuum expectation value of two Wilson
links in the directions  $\bar\xi_{J_i}=\left(1,{\hat n}_{J_i}\right)/\sqrt{2}$
with ${\hat n}_{J_i}={\bf P}_{J_i}/|{\bf P}_{J_i}|$, $i=1$, 2. An explicit
next-to-leading-order (NLO) calculation in the Mellin ($N$) space gives
\begin{eqnarray}
S^{(1)}=
\frac{\alpha_s C_F}{\pi(R E_{J_H})^2}\ln\frac
{\bar\xi_{J_1}^2\bar\xi_{J_2}^2}
{4(\bar\xi_{J_1}\cdot \bar\xi_{J_2})^2}\left(\frac{1}{\epsilon}
+\ln\frac{4\pi\mu_f^2\bar N^2}{R^4E_{J_H}^2e^{\gamma_E}}\right),
\end{eqnarray}
where the color factor $C_F=4/3$ and the moment $\bar N\equiv
N\exp(\gamma_E)$, with $\gamma_E$ being the Euler constant.
The off-shellness $\bar\xi_{J_i}^2$ associated with the
$b$-quark jets implies that $S^{(1)}$
contains the collinear dynamics which has been absorbed into the jet functions.
Hence, the subtraction of the collinear divergences from the soft function
is necessary to avoid double counting.
The collinear divergences from loop momenta collimated to the $b$ quark
($\bar b$ quark) can be collected with the $\bar b$ quark ($b$ quark) line
being replaced by the eikonal line in the direction $n_{J_1}$ ($n_{J_2}$)
that appear in the $b$-quark jet definitions \cite{Almeida:2008tp}. The NLO subtraction term
for the latter with the same cone radius $R$ is obtained by substituting the vector
$n_{J_2}$ for $\bar\xi_{J_1}$ in $S^{(1)}$.
After this subtraction, we have
\begin{eqnarray}
S^{(1)}-S^{(1)}_{n_{J_2}}=\frac{\alpha_s C_F}{\pi(R E_{J_H})^2}\ln\frac
{\bar\xi_{J_1}^2(\bar\xi_{J_2}\cdot n_{J_2})^2}
{(\bar\xi_{J_1}\cdot \bar\xi_{J_2})^2n_{J_2}^2}\left(\frac{1}{\epsilon}
+\ln\frac{4\pi\mu_f^2\bar N^2}{R^4E_{J_H}^2e^{\gamma_E}}\right),
\end{eqnarray}
to which we can further impose the condition $4(\bar\xi_{J_2}\cdot n_{J_2})^2/n_{J_2}^2=R^2$
for defining a quark (or gluon) jet \cite{Li:2011hy}.
Because the thin jet $J_1$ contributes only the overall normalization
in ${\bar J}_{H}^E$, the choice of $n_{J_1}$ is arbitrary.
We then utilize this freedom, and
choose $n_{J_1}$ such that $S^{(1)}_{n_{J_1}}$ has the logarithmic coefficient
the same as of $S^{(1)}-S^{(1)}_{n_{J_2}}$.
This choice is possible, because of $\bar\xi_{J_1}\cdot \bar\xi_{J_2}\sim (m_H/E_{J_H})^2
\sim O(r)$ in the considered kinematic region. The further collinear subtraction
leads to
\begin{eqnarray}
S^{(1)}-S^{(1)}_{n_{J_1}}-S^{(1)}_{n_{J_2}}\approx 0,
\end{eqnarray}
so the soft function in this special scheme
is given by $S(\omega,R,\mu_f^2)\approx\delta(\omega)$.

Equation~(\ref{he0}) then reduces to
\begin{eqnarray}
{\bar J}_{H}^E(1,E_{J_H},R,r)& =&\frac{1}{R^2 (E_{J_H})^3}
\frac{1}{\pi^2}\left(\frac{m_b}{v}\right)^2\int dE_{J_1}
\int d\cos\theta_{J_1} (E_{J_1}\cos\theta_{J_1}+E_{J_2}\cos\theta_{J_2})
\nonumber\\
& &\times\left[E_{J_1}^2\Theta(ar-\theta_{J_1})+R^2E_{J_2}^3{\bar J}_2^E(1,E_{J_2},R,r)
\Theta(ar-\theta_{J_2})\right]\nonumber\\
& &\times
\frac{E_{J_1}E_{J_2}[1-\cos(\theta_{J_1}+\theta_{J_2})]}
{\{2E_{J_1}E_{J_2}[1-\cos(\theta_{J_1}+\theta_{J_2})]
-m_{H}^2\}^2+\Gamma_H^2 m_{H}^2},\label{final}
\end{eqnarray}
where the light-quark jet
functions are set at the factorization scale $\mu_f^2=(E_{J}R)^2/\bar N$,
the renormalization scale for ${\bar J}_{H}^E$ is chosen to be
$\mu=E_{J_H}r/R$ \cite{Li:2011hy}, and the Mellin transformation
${\bar J}_{2}^E(1,E_{J_H},R,r)\equiv
\int J_{2}^E(m_{J_2}^2,E_{J_H},R,r) d m_{J_2}^2/(R E_{J_2})^2$ has
been inserted.

The choice of the merging parameter $a$ is a matter of
factorization schemes, and the difference arising from distinct $a$'s
will be compensated by the corresponding distinct hard kernels $H^E$.
That is, a larger $a$ means more contribution to the Higgs JEP from the
$b$-quark jets, and less contribution from $H^E$. Since we neglect $H^E$
and consider only the LO hard kernel $H^{(0)}$ here,
our analysis will be more consistent, as
a larger $a$ is chosen. Below, we set $a$ in Eq.~(\ref{final}) to its maximal
allowed value, $a=2$, according to the prescription of the cone algorithm,
and predict the Higgs JEP with $E_{J_H}=500$ GeV and $E_{J_H}=1000$ GeV,
both with a cone radius of $R=0.7$.
A JEP is defined as
\begin{eqnarray}
\Psi(E_{J},R,r) = \frac{{\bar J}^E(1,E_J,R,r)}{{\bar J}^E(1,E_J,R,R)}.
 \label{eqn:psi}
\end{eqnarray}
It is interesting to note that a simple expression can be
derived for the JEP of a boosted Higgs jet after applying
the narrow width approximation for the Higgs boson propagator.
It yields
\begin{eqnarray}\label{app}
\Psi(E_{J_H},R,r)=\frac{\int_{z_{m}(r)}^1dz z(1-z)[1+\Psi_q(zE_{J_H},R,r)]}
{\int_{z_{m}(R)}^1dz z(1-z)[1+\Psi_q(zE_{J_H},R,R)]},
\end{eqnarray}
where the integration variable $z=E_{J_1}/E_{J_H}$, the lower limit
$z_{m}(r)={\hat m}_H^2/({\hat m}_H^2+a^2r^2)$, the small parameter
$\hat m_{H}\equiv m_{H}/E_{J_H}$, and $\Psi_q$ denotes the light-quark
JEP~\cite{Li:2011hy,Li:2012bw}.
Compared to the energy profiles of QCD
jets~\cite{Li:2011hy},
the Higgs JEP is lower at small $r$ due to the dead-cone
effect, and increases faster with $r$ once the energetic $b$-quark jets
start to contribute.

\begin{figure}[h]
\centering
\begin{minipage}{0.45\textwidth}
\includegraphics[width=\textwidth]{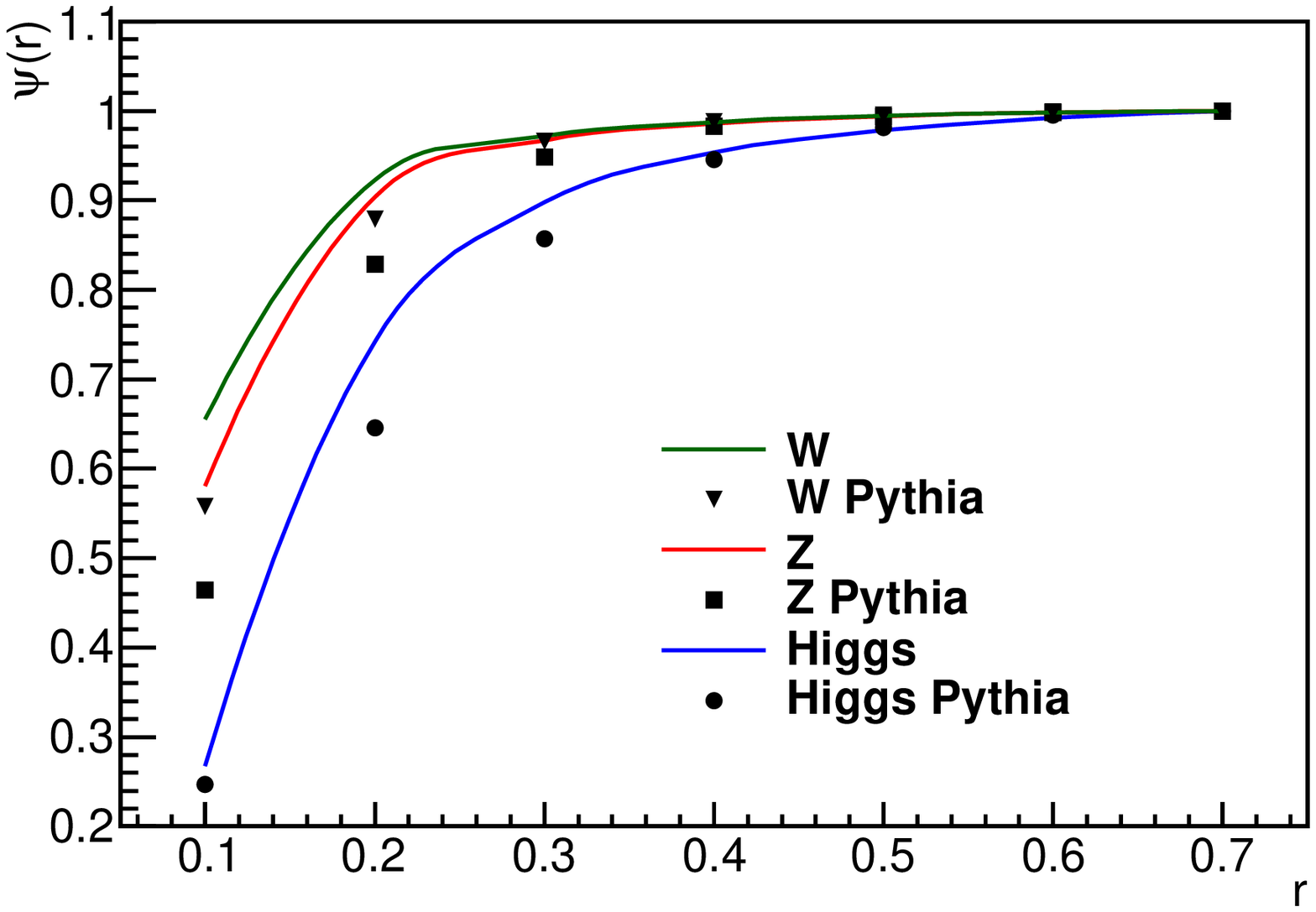}
 \caption{Comparison of $W$, $Z$ and Higgs JEPs to Pythia8 predictions,
 for $E_J=500$ GeV and $R=0.7$.}
 \label{fig:Pythia}
\end{minipage}
\hfill
\begin{minipage}{0.45\textwidth}
   \includegraphics[width=\textwidth]{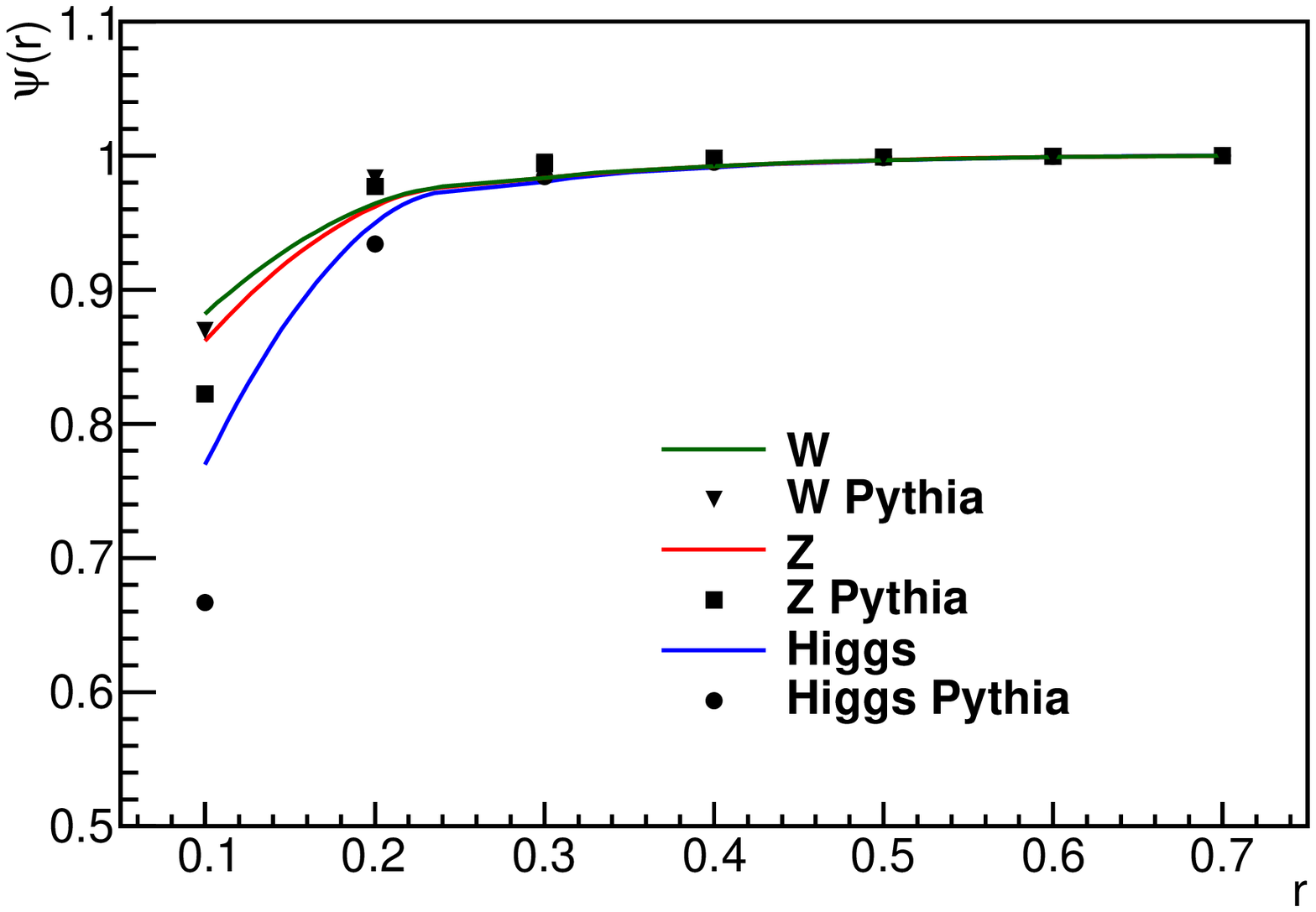}
 \caption{Comparison of $W$, $Z$ and Higgs JEPs to Pythia8 predictions,
 for $E_J=1000$ GeV and $R=0.7$.}
 \label{fig:Pythia1000}
\end{minipage}
\end{figure}

\begin{figure}[h]
\centering
  \includegraphics[scale=0.4]{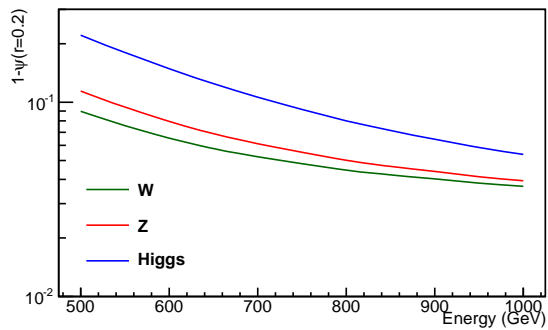}
 \caption{Energy dependence of the JEPs for Higgs, $W$, and $Z$ for fixed $r=0.2$ and $R=0.7$.}
 \label{fig:FixedR}
\end{figure}

Our formalism can be readily extended for studying
boosted $W$ and $Z$ bosons in their hadronic decay modes,
by inserting their masses and widths, since Eq.~(\ref{app}) is coupling-
and spin-independent. As shown in Fig.~\ref{fig:Pythia},
the predicted $W$, $Z$ and Higgs JEPs are well consistent with those from
Pythia8 \cite{Sjostrand:2007gs} for the heavy-boson
jet energy $E_J=500$ GeV. For the Pythia8 comparison, we have
used the 4C tune which was shown to agree well with the ATLAS data for the JEPs of
QCD jets with energies ranging from 30 GeV to 600 GeV \cite{Aad:2011kq}.
In the analysis we include the effects of initial-state and
final-state radiations, hadronization, and
beam remnants, but turn off the effects from multiple parton interactions.
Similar agreement is also observed, cf. Fig.~\ref{fig:Pythia1000}, 
for these boosted electroweak bosons at 1 TeV, though the deviation at $r=0.1$ is larger.
From these figures, we can get a rough estimate of the effective radii needed
to capture all of the radiation for these boosted electroweak bosons. We find that
the effective radius needed is $R\approx 0.7$ and $R\approx 0.3$ for the Higgs
boson at 500 GeV and 1 TeV, respectively, and 0.4 and 0.25 for either $W$ or $Z$ bosons.
Moreover, the $W$ and $Z$ JEPs are thinner than the Higgs JEP due to
their smaller masses. This is further demonstrated via the energy dependence of the JEP
for a fixed $r$ value ($r=0.2$) in Fig.~\ref{fig:FixedR}: the separation of the $W$, $Z$,
and Higgs jets becomes more difficult as the jet energy increases,
because the ratio of the mass to the jet energy becomes smaller.
Note that the simple expression in Eq.~(\ref{app})
is derived under the aforementioned approximations, at the LO
accuracy for the hard kernels, and with the neglect of soft links
among the heavy-boson jet and other subprocesses, such as beams and
other final-state particles. The associated theoretical uncertainties
can be reduced by taking into account relevant corrections, and will be addressed in a future work.
Furthermore, the questions of how techniques like trimming \cite{Krohn:2009th}, pruning \cite{Ellis:2009me,Ellis:2009su}, soft-drop \cite{Dasgupta:2013yea,Dasgupta:2013ihk},
and other similar techniques affect the JEP, how subleading corrections
affect the JEP, and what happens to the profile if we loosen some of the
aforementioned approximations will be investigated in a future work.

In conclusion, we have applied the pQCD factorization to
formulate the JEP of a boosted colorless heavy-particle
(such as $W$, $Z$, Higgs, $W^\prime$ and $Z^\prime$ boson)
jet, which is found to differ dramatically from the ordinary QCD JEPs
with similar energy and jet radius. The formalism is greatly simplified
by considering inside the boosted jet a thin jet and a fat jet, which
absorbs all the soft-gluon effect, when the heavy-particle decays into a
quark and antiquark pair.
More interestingly, the analytical expression for the JEPs
of the electroweak bosons allows for an
easy and efficient discrimination of different production processes
for these boosted jets.
The implementation of this discrimination method can further help
suppress QCD background to signals of boosted $W$ and $Z$ jets,
after applying conventional kinematic selections.

\section*{Acknowledgements}
This work was supported by Ministry of Science and Technology of R.O.C. under
Grant No. NSC-101-2112-M-001-006-MY3;
by the Natural Science Foundation of China under Grant NO. 11305179;
by Youth Innovation Promotion Association, CAS; and
by the U.S. National Science Foundation under Grand No. PHY-1417326.


\section*{References}
\bibliographystyle{elsarticle-num}
\bibliography{higgsjets}

\begin{thebibliography}{10}
\expandafter\ifx\csname url\endcsname\relax
  \def\url#1{\texttt{#1}}\fi
\expandafter\ifx\csname urlprefix\endcsname\relax\def\urlprefix{URL }\fi
\expandafter\ifx\csname href\endcsname\relax
  \def\href#1#2{#2} \def\path#1{#1}\fi

\bibitem{deFlorian:2013jea}
D.~de~Florian, J.~Mazzitelli, {Higgs Boson Pair Production at
  Next-to-Next-to-Leading Order in QCD}, Phys.Rev.Lett. 111 (2013) 201801.
\newblock \href {http://arxiv.org/abs/1309.6594} {\path{arXiv:1309.6594}},
  \href {http://dx.doi.org/10.1103/PhysRevLett.111.201801}
  {\path{doi:10.1103/PhysRevLett.111.201801}}.

\bibitem{Butterworth:2008iy}
J.~M. Butterworth, A.~R. Davison, M.~Rubin, G.~P. Salam, {Jet substructure as a
  new Higgs search channel at the LHC}, Phys.Rev.Lett. 100 (2008) 242001.
\newblock \href {http://arxiv.org/abs/0802.2470} {\path{arXiv:0802.2470}},
  \href {http://dx.doi.org/10.1103/PhysRevLett.100.242001}
  {\path{doi:10.1103/PhysRevLett.100.242001}}.

\bibitem{Almeida:2008yp}
L.~G. Almeida, S.~J. Lee, G.~Perez, G.~F. Sterman, I.~Sung, et~al.,
  {Substructure of high-$p_T$ Jets at the LHC}, Phys.Rev. D79 (2009) 074017.
\newblock \href {http://arxiv.org/abs/0807.0234} {\path{arXiv:0807.0234}},
  \href {http://dx.doi.org/10.1103/PhysRevD.79.074017}
  {\path{doi:10.1103/PhysRevD.79.074017}}.

\bibitem{Almeida:2010pa}
L.~G. Almeida, S.~J. Lee, G.~Perez, G.~Sterman, I.~Sung, {Template Overlap
  Method for Massive Jets}, Phys.Rev. D82 (2010) 054034.
\newblock \href {http://arxiv.org/abs/1006.2035} {\path{arXiv:1006.2035}},
  \href {http://dx.doi.org/10.1103/PhysRevD.82.054034}
  {\path{doi:10.1103/PhysRevD.82.054034}}.

\bibitem{Thaler:2010tr}
J.~Thaler, K.~Van~Tilburg, {Identifying Boosted Objects with N-subjettiness},
  JHEP 1103 (2011) 015.
\newblock \href {http://arxiv.org/abs/1011.2268} {\path{arXiv:1011.2268}},
  \href {http://dx.doi.org/10.1007/JHEP03(2011)015}
  {\path{doi:10.1007/JHEP03(2011)015}}.

\bibitem{Almeida:2008tp}
L.~G. Almeida, S.~J. Lee, G.~Perez, I.~Sung, J.~Virzi, {Top Jets at the LHC},
  Phys. Rev. D79 (2009) 074012.
\newblock \href {http://dx.doi.org/10.1103/PhysRevD.79.074012}
  {\path{doi:10.1103/PhysRevD.79.074012}}.

\bibitem{Li:2011hy}
H.-n. Li, Z.~Li, C.-P. Yuan, {QCD resummation for jet substructures}, Phys.
  Rev. Lett. 107 (2011) 152001.
\newblock \href {http://arxiv.org/abs/1107.4535} {\path{arXiv:1107.4535}}.

\bibitem{Ellis:2010rwa}
S.~D. Ellis, C.~K. Vermilion, J.~R. Walsh, A.~Hornig, C.~Lee, {Jet Shapes and
  Jet Algorithms in SCET}, JHEP 1011 (2010) 101.
\newblock \href {http://dx.doi.org/10.1007/JHEP11(2010)101}
  {\path{doi:10.1007/JHEP11(2010)101}}.

\bibitem{Kelley:2011tj}
R.~Kelley, M.~D. Schwartz, H.~X. Zhu, {Resummation of jet mass with a jet
  veto}\href {http://arxiv.org/abs/1102.0561} {\path{arXiv:1102.0561}}.

\bibitem{Li:2012bw}
H.-n. Li, Z.~Li, C.~P. Yuan, {QCD resummation for light-particle jets}, Phys.
  Rev. D87 (2013) 074025.
\newblock \href {http://arxiv.org/abs/1206.1344} {\path{arXiv:1206.1344}},
  \href {http://dx.doi.org/10.1103/PhysRevD.87.074025}
  {\path{doi:10.1103/PhysRevD.87.074025}}.

\bibitem{Sjostrand:2007gs}
T.~Sjostrand, S.~Mrenna, P.~Z. Skands, {A Brief Introduction to PYTHIA 8.1},
  Comput. Phys. Commun. 178 (2008) 852--867.
\newblock \href {http://dx.doi.org/10.1016/j.cpc.2008.01.036}
  {\path{doi:10.1016/j.cpc.2008.01.036}}.

\bibitem{Aad:2011kq}
G.~Aad, et~al., {Study of Jet Shapes in Inclusive Jet Production in $pp$
  Collisions at $\sqrt{s}=7$ TeV using the ATLAS Detector}, Phys. Rev. D83
  (2011) 052003.
\newblock \href {http://arxiv.org/abs/1101.0070} {\path{arXiv:1101.0070}},
  \href {http://dx.doi.org/10.1103/PhysRevD.83.052003}
  {\path{doi:10.1103/PhysRevD.83.052003}}.

\bibitem{Krohn:2009th}
D.~Krohn, J.~Thaler, L.-T. Wang, {Jet Trimming}, JHEP 1002 (2010) 084.
\newblock \href {http://dx.doi.org/10.1007/JHEP02(2010)084}
  {\path{doi:10.1007/JHEP02(2010)084}}.

\bibitem{Ellis:2009me}
S.~D. Ellis, C.~K. Vermilion, J.~R. Walsh, {Recombination Algorithms and Jet
  Substructure: Pruning as a Tool for Heavy Particle Searches}, Phys. Rev. D81
  (2010) 094023.
\newblock \href {http://arxiv.org/abs/0912.0033} {\path{arXiv:0912.0033}},
  \href {http://dx.doi.org/10.1103/PhysRevD.81.094023}
  {\path{doi:10.1103/PhysRevD.81.094023}}.

\bibitem{Ellis:2009su}
S.~D. Ellis, C.~K. Vermilion, J.~R. Walsh, {Techniques for improved heavy
  particle searches with jet substructure}, Phys. Rev. D80 (2009) 051501.
\newblock \href {http://dx.doi.org/10.1103/PhysRevD.80.051501}
  {\path{doi:10.1103/PhysRevD.80.051501}}.

\bibitem{Dasgupta:2013yea}
M.~Dasgupta, S.~Marzani, G.~P. Salam, {QCD calculations for jet
  substructure}[Nuovo Cim.C037,no.02,131(2014)].
\newblock \href {http://arxiv.org/abs/1311.6514} {\path{arXiv:1311.6514}},
  \href {http://dx.doi.org/10.1393/ncc/i2014-11746-x}
  {\path{doi:10.1393/ncc/i2014-11746-x}}.

\bibitem{Dasgupta:2013ihk}
M.~Dasgupta, A.~Fregoso, S.~Marzani, G.~P. Salam, {Towards an understanding of
  jet substructure}, JHEP 09 (2013) 029.
\newblock \href {http://arxiv.org/abs/1307.0007} {\path{arXiv:1307.0007}},
  \href {http://dx.doi.org/10.1007/JHEP09(2013)029}
  {\path{doi:10.1007/JHEP09(2013)029}}.

\end{thebibliography}

\end{document}